\documentclass[preprint]{aastex}
\usepackage{epsf}

\def\plotone#1#2{\centering \leavevmode \epsfxsize=#1\columnwidth \epsfbox{#2}}

\shorttitle{???}
\shortauthors{Schubert and Zhang}

\begin{document}


\title{Spherical, Oscillatory $\mbox{\boldmath$\alpha^2$}$-Dynamo Induced by Magnetic Coupling \\
    Between a Fluid Shell and an Inner Electrically Conducting Core:\\
    Relevance to the Solar Dynamo}


\author{G. Schubert}
\affil{Department of Earth and Space Sciences,\\
 Institute of Geophysics and Planetary Physics,\\
 University of California,
    Los Angeles, CA 90095}
\email{schubert@ucla.edu}

\and

\author{K. Zhang}
\affil{School of Mathematical Sciences\\
University of Exeter, Exeter EX4 4QE\\
United Kingdom}
\email{KZhang@maths.ex.ac.uk}





\begin{abstract}
A two-layer spherical $\alpha^2$-dynamo model  consisting of an inner electrically conducting core (magnetic diffusivity $\lambda_i$ and radius
$r_i$) with $\alpha = 0$ surrounded by an electrically conducting spherical shell (magnetic diffusivity $\lambda_o$ and radius $r_o$) with
a constant $\alpha$ is shown to exhibit oscillatory behavior for values of $\beta = \lambda_i/\lambda_o$ and $r_i/r_o$  relevant to the solar dynamo.
Time-dependent dynamo solutions require $r_i/r_o \geq 0.55$  and  $\beta \leq O(1)$. For the Sun,  $r_i/r_o$ is about 0.8  and $\beta\approx 10^{-3}$. The time scale  of the oscillations matches the 22   year period of the
sunspot cycle for  $\lambda_0 = O(10^2\mbox{\thinspace km}^2\mbox{\thinspace s}^{-1}$). It is unnecessary to hypothesize an $\alpha\omega$-dynamo to obtain  oscillatory dynamo solutions; an $\alpha^2$-dynamo suffices provided the spherical
shell region of dynamo action lies above a large, less magnetically diffusive core, as is the case for the solar dynamo.
\end{abstract}




\section{Introduction}

$\alpha^2$- and $\alpha\omega$-dynamo models have
been studied for decades
\citep{Bra64,Ste66,Rob72,Mof78,Gub87,Bar87}.
The $\alpha^2$-dynamo is usually stationary
\citep[e.g.,][]{Rob72,Gub87,Hol96},
although oscillatory $\alpha^2$-dynamos have been found to occur
in the special circumstance wherein $\alpha$ changes rapidly in
boundary layers
\citep{Rad87,Bar87}.
In such cases, the period of the $\alpha^2$-dynamo depends strongly
on the location of the $\alpha$-boundary layer and is typically
an order of magnitude or more smaller than the magnetic diffusion
time across the dynamo generation region. In general, oscillatory
dynamo behavior has been produced by combination of the
$\alpha$- and $\omega$-effects.  Kinematic models of the solar dynamo, which is 
inherently oscillatory, have been of the $\alpha\omega$-type
\citep[e.g.,][]{Roa97}.

In this paper, we report that spherical oscillatory $\alpha^2$-dynamos
can be simply induced by the magnetic coupling between an
electrically conducting outer fluid shell and a conducting
inner spherical core even when $\alpha$ in the outer shell is
a constant.  The period of oscillation is of the same order
of magnitude as the magnetic diffusion time across the outer
shell and depends largely on the electrical conductivity
of the inner core.  Oscillatory behavior occurs when
the outer region of dynamo action surrounds a large, less
magnetically diffusive core.
The radiative interior of the Sun
is a large region with a smaller magnetic diffusivity than
the overlying convection zone wherein dynamo action occurs.
The oscillatory character of the Sun's magnetic field, as expressed in
the 22 year periodicity of the sunspot cycle, could then be related
to the electromagnetic coupling of the region of magnetic field generation
in the convection zone with the radiative core of the Sun. The importance
of an inner electrically conducting core to the problem of magnetic
field generation in an overlying spherical shell has been emphasized
in the $\alpha^2$-type models of the geodynamo by 
\citet{Hol93,Hol95},
\citet{Hol96},
and
\cite{Gub99}.
Nevertheless, the effects of an inner core, with magnetic diffusivity
different from that of the overlying fluid convecting shell in
which dynamo action takes place, have not been fully elucidated.

The problem of inner core-fluid shell coupling is made
difficult by complicated electromagnetic matching conditions at
the interface between the regions. For this reason, and to facilitate understanding
of the physical effects, we consider the simplest type of 
$\alpha^2$-dynamo model consisting of a spherical shell with
$\alpha = \mbox{constant}$ surrounding a core with $\alpha = 0$.
We derive the appropriate matching conditions for a core and shell
of arbitrary magnetic diffusivity. These matching conditions do not
appear to have been considered in previous studies of spherical
$\alpha^2$-dynamos and they result in oscillatory dynamo solutions.

\section{Model, Equations and Boundary Conditions}

The model consists of a turbulent fluid spherical shell of inner radius
$r_i$ and outer radius $r_o$ with constant (turbulent) magnetic diffusivity
$\lambda_o$.  A magnetic field is generated in the shell by the
$\alpha$-effect
\citep{Ste66,Rob72}.
For  $r > r_o$, we assume there is a non-conductor; for
$r < r_i$ we assume that there is a conductor with magnetic
diffusivity $\lambda_i$. The kinematics of the $\alpha^2$-dynamo
in the spherical shell is governed by the non-dimensional linear
equations for the magnetic field  $\mathbf{B}_o$
\begin{equation}
  \frac{\partial\mathbf{B}_o}{\partial t}  = R\left(1-\eta\right)
                                             \nabla\times\alpha\mathbf{B}_o
                                             +\nabla^2\mathbf{B}_o
\end{equation}
\begin{equation}
   \nabla\cdot \mathbf{B}_o = 0
\end{equation}
In the inner sphere the magnetic field $\mathbf{B}_i$ is governed by
\begin{equation}
   \frac{\partial\mathbf{B}_i}{\partial t} = \beta\nabla^2\mathbf{B}_i
\end{equation}
\begin{equation}
   \nabla \cdot \mathbf{B}_i = 0
\end{equation}
Equations (1)--(4) are scaled by the thickness of the shell
$\left(r_o - r_i\right)$ and by the magnetic diffusion timescale
$\left(r_o - r_i\right)^2/\lambda_o$. The scaling of the linear system
of equations for the magnetic field is arbitrary. The non-dimensional
parameters in the above equations, $\beta$, $\eta$, and the
magnetic Reynolds number $R$ are defined as
\begin{equation}
 \beta = \frac{\lambda_i}{\lambda_o},\quad \eta = \frac{r_i}{r_o},\quad
          R = \frac{r_o\alpha}{\lambda_o}
\end{equation}
Since the main purpose of this paper is to understand the effect
of an electrically conducting inner core, we adopt the simplest possible
model and take $\alpha$ constant in the spherical shell
$r_i < r < r_o$; $\alpha$ is zero outside the shell.
With this assumption, spherical harmonics are decoupled and the problem
is reduced to a one-dimensional problem with complicated boundary conditions.

At the interface between the shell and the perfectly insulating exterior,
i.e., at $r = r_o$, the magnetic field must be continuous
\begin{equation}
  \mathbf{B}_o = \mathbf{B}^{(e)}\quad \mbox{at}\quad r = r_o
\end{equation}
where $\mathbf{B}^{(e)} = -\nabla\phi$ is the magnetic field
in the insulating exterior
$r > r_o$, and $\nabla^2\phi = 0$. At the interface
between the shell and the conducting inner sphere, i.e., 
at $r = r_i$, both the magnetic field $\mathbf{B}$
and tangential components of the electric field
$\mathbf{E}$ must be continuous
\begin{equation}
 {\bf B}_o = \mathbf{B}_i,\quad \mbox{\boldmath$\hat r$} \times
                  \mathbf{E}_o = \mbox{\boldmath$\hat r$} \times
                  \mathbf{E}_i\quad \mbox{at}\quad  r = r_i
\end{equation}
where $\mbox{\boldmath$\hat r$}$ is the unit radial vector,
$\mathbf{E}_o$ is the electric field in the outer shell, and
$\mathbf{E}_i$ is the electric field in the inner core.

Conditions (2) and (4) allow us to express the magnetic fields as
a sum of poloidal and toroidal vectors
\begin{eqnarray}
   \mbox{\boldmath{$B$}}_o &=& \nabla \times \nabla \times \mbox{\boldmath{$r$}}h_o + \nabla
                     \times \mbox{\boldmath{$r$}}g_o\\
   \mbox{\boldmath{$B$}}_i &=& \nabla \times \nabla \times \mbox{\boldmath{$r$}}h_i + \nabla
                     \times \mbox{\boldmath{$r$}}g_i
\end{eqnarray}
where $\mbox{\boldmath$r$}$ is the position vector.  Use of Equation (8) in boundary condition
(6) and expansion of $h_o$ and $g_o$ in terms of spherical harmonics give
\begin{equation}
  g_o = 0,\quad \frac{\partial h_o}{\partial r} + \frac{(l + 1) h_o}{r}
    = 0\quad \mbox{at}\ r = r_o
\end{equation}
where $l$ is the degree of the spherical harmonic $Y_l^m$.

Extra care must be taken for the magnetic boundary conditions
at the interface $r = r_i$.  There are four different cases that we
have studied:

\noindent (I). The limit $\beta\rightarrow \infty$ for both
 stationary and oscillatory dynamos.  In this case, the boundary
condition for the magnetic field is simply
\begin{equation}
   g_o = 0,\quad \frac{\partial h_o}{\partial r} 
          - \frac{lh_o}{r} = 0\quad\mbox{at}\ r = r_i
\end{equation}
\noindent(II).  The limit $\beta\rightarrow 0$ for a stationary dynamo.
In this case, boundary conditions (7) require
\begin{equation}
  h_o = 0,\quad R \left(1-\eta\right)\, r\,\frac{\partial h_o}{\partial r}
         - \frac{\partial (rg_o)}{\partial r} = 0\quad
         \mbox{at}\quad  r = r_i
\end{equation}
\noindent(III). The limit $\beta\rightarrow 0$ for an oscillatory dynamo.
In this case, boundary conditions (7) require
\begin{equation}
  \frac{\partial h_o}{\partial r} -\frac{l h_o}{r} = 0,\quad
              R\left(1-\eta\right) \left(l + 1\right) h_o
               -\frac{\partial (rg_o)}{\partial r} = 0
              \qquad \mbox{at}\  r = r_i
\end{equation}
\noindent (IV). The general case for $\beta$ not tending toward 
0 or $\infty$, for both stationary and oscillatory dynamos.
In this case, boundary conditions (7) require
\begin{equation}
 g_o = g_i,\quad h_o = h_i,\quad \frac{\partial h_o}{\partial r}
      = \frac{\partial h_i}{\partial r},\quad  R\left(1-\eta\right)\,
        \frac{\partial (rh_o)}{\partial r} -\frac{\partial(rg_o)}{\partial r}
         + \beta\,\frac{\partial (rg_i)}{\partial r} = 0
         \quad \mbox{at}\quad  r = r_i
\end{equation}
The last case is evidently the most complicated one. The solutions
presented below show that for $\beta \geq 10$, case I provides
a good approximation to case IV, while for $\beta \leq 0.1$, cases
II and III provide a good approximation to case IV. In cases II through
IV, $g$ and $h$ are coupled by boundary conditions (7). The solutions
are invariant to a change in the sign of $R$.

\section{Solution Method}

In all cases, solutions are expanded in terms of spherical harmonics,
implicit in the forms of the boundary and interface conditions
given above. The spherical harmonics are decoupled and only the lowest one
$\left(\ell = 1\right)$ is used in the analysis. The $\ell = 2$ mode,
not discussed here, behaves similarly to the $\ell = 1$ mode.

The time dependence of the solutions is written as
$\exp \left(\sigma_r + i\omega t\right)$ and onset of dynamo action
$\left(\sigma_r = 0\right)$ is sought.  As discussed below,
dynamos are either stationary $\left(\omega = 0\right)$ or
oscillatory $\left(\omega\neq 0\right)$. The frequency $\omega$
is dimensionless with respect to the timescale
$\left(r_o - r_i\right)^2/\lambda_o$.

In case I, a perfectly insulating core, and in cases II and III,
a perfectly conducting core, it is only necessary to
solve for $g_o$ and $h_o$ subject to the above boundary conditions
at $r_i$ and $r_o$. Exact analytic solutions for $g_o$ and $h_o$
in these cases can be found in terms of the spherical Bessel
functions of the first and second kind.  The solutions reduce
to finding the eigenvalues of a $4\times 4$ matrix. The eigenvalues
give critical values of the magnetic Reynolds number $R$
as a function of $\eta = r_i/r_o$, for which steady or
oscillatory dynamos are possible (i.e., $\sigma_r = 0$).

In case IV, a core of arbitrary $\beta$, solutions must be obtained
in both the shell and the core subject to the above matching conditions,
i.e., $g_o$, $h_o$, $g_i$, and $h_i$ must be determined.
In principle, analytic solutions are possible, but it is computationally
more efficient to seek numerical solutions. We do this by employing
a spectral-Tau method which expands solutions in terms of Chebyshev
polynomials. The numerical solutions for arbitrary $\beta$, and the
analytic solutions determined independently in cases I, II, and III,
provide a mutual validation of the separate methods. For appropriate
values of $\beta$, the solutions of the separate methods agree
essentially exactly.

\section{Results}

The principal results of this study are summarized in Figure~1
\begin{figure}[b!]
\plotone{.6}{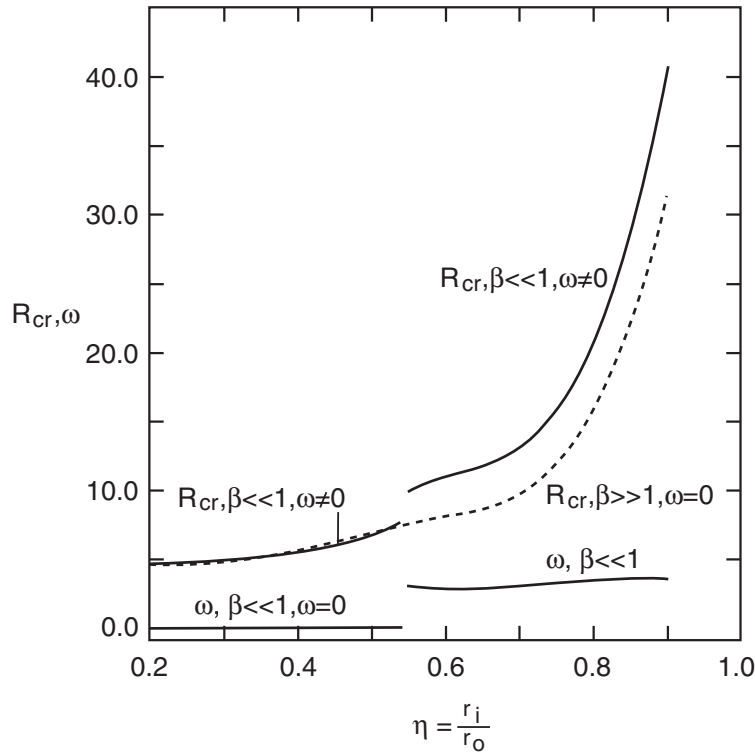}
\caption{Critical magnetic Reynolds number $R_{cr}$
and dimensionless frequency $\omega$ of dipolar dynamo solutions
at the onset of dynamo action vs.\ $\eta$. \label{fig1}}
\end{figure}
which gives the critical magnetic Reynolds number $R_{cr}$
for the onset of dynamo action in the dipole $(l = 1)$ mode
as a function of $\eta$, the ratio of the inner radius
of the shell to its outer radius.  The critical value of $R$
increases with increasing $\eta$.  For an insulating core
($\beta > > 1$, dashed curve) dynamo solutions are always steady
$\left(\omega\equiv 0\right)$.  For a perfectly conducting core
($\beta < < 1$, solid curve) there are two branches of dynamo
solutions depending on $\eta$; for $\eta\leq 0.55$
the dynamo is steady, but for $\eta \geq 0.55$ the dynamo
is oscillatory.  There is a jump in $R_{cr}$ at the transition
from steady dynamos to oscillatory dynamos near $\eta = 0.55$.
The dimensionless frequency $\omega$ of the oscillatory dynamos,
also shown in Figure 1 as a function of $\eta$, varies between
about 2.5 and 3 for all values of $\eta$ considered.

The values of $R_{cr}$ for arbitrary $\beta$ lie in the narrow space
between the solid and dashed curves of Figure 1.  Importantly,
it is found that $\beta$ need not in fact be very small compared with
unity for oscillatory dynamos to exist.  For example, when
$\eta = 0.8$, the value appropriate to the solar dynamo,
oscillatory dynamo solutions are found for $\beta$ less
than 2 to 3.

\section{Discussion}

The problem solved above is a classically simple one of the type
considered by 
\citet{Ste66}
and 
\citet{Rob72}
decades ago.  Yet the effects
of an inner electrically conducting core on the
$\alpha^2$-dynamo are subtle and not heretofore appreciated.
They enter through the complicated electromagnetic matching conditions
at the interface between the core and the surrounding shell in which
dynamo action occurs. The main effect of the core is to introduce
time dependence into the dynamo solutions for cores whose radii
are greater than about 0.55 of the outer radius of the shell.
An additional requirement for time dependence is that the core
be a reasonably good electrical conductor; in terms of the magnetic
diffusivity ratio $\beta = \lambda_i/\lambda_o$,
$\beta\leq O(1)$ suffices for oscillatory dynamo behavior.

The importance of all this to the solar dynamo is that the parameters
of the solar dynamo satisfy the requirements of oscillatory
$\alpha^2$-dynamo solutions.  For the solar dynamo
$\eta$ is about 0.8 and $\beta$ is about $10^{-3}$
\citep{Mof78}.
In addition, if $\omega$ (from Figure 1) is made dimensional using
the time scale $\left(r_o - r_i\right)^2/\lambda_o$ with
$r_o - r_i = 1.4 \times 10^5~\mbox{km}$ and
$\lambda_o = O\left(10^2\mbox{\thinspace km}^2\mbox{\thinspace s}^{-1}\right)$
(eddy magnetic diffusivity), then the period of the
oscillatory dynamo solution is comparable to the 22 year
period of the sunspot cycle. Thus, $\alpha^2$-dynamo action
alone could be responsible for the observed time dependence
of the large scale solar magnetic field. 
It is not our intent to suggest that the $\omega$-effect
is not significant in dynamo action in general, or in the solar
dynamo in particular, because it represents a physically important
process.  Our purpose is only to clarify some physics and demonstrate
the potential importance of a hitherto overlooked effect, that of
the oscillatory $\alpha^2$-dynamo.  A detailed analysis of the
cases $\alpha = \alpha\/(r)$ and $\alpha\propto \cos \theta
(\theta = \mbox{polar\ angle})$ is in progress.

\acknowledgments

G.\ S.\ acknowledges support from NASA's Planetary Atmospheres Program.
K.\ Z.\ is supported by PPARC and NATO grants.





\clearpage





\end{document}